\documentclass[preprint,aps,floats,subeqn,showpacs]{revtex4}
\usepackage{epsfig}

\newcommand{\be}{\begin{equation}}
\newcommand{\ee}{\end{equation}}
\newcommand{\bea}{\begin{eqnarray}}
\newcommand{\eea}{\end{eqnarray}}
\newcommand{\bml}{\begin{mathletters}}
\newcommand{\eml}{\end{mathletters}}

\begin{document}

\tighten

\draft




\title{Inflating branes inside abelian strings}
\renewcommand{\thefootnote}{\fnsymbol{footnote}}
\author{Yves Brihaye\footnote{yves.brihaye@umh.ac.be}}
\affiliation{Facult\'e des Sciences, Universit\'e de Mons-Hainaut,
7000 Mons, Belgium}
\author{Terence Delsate}
\affiliation{Facult\'e des Sciences, Universit\'e de Mons-Hainaut,
7000 Mons, Belgium}
\author{Betti Hartmann\footnote{b.hartmann@iu-bremen.de}}
\affiliation{School of Engineering and Science, International University Bremen (IUB),
28725 Bremen, Germany}

\date{\today}
\setlength{\footnotesep}{0.5\footnotesep}

\begin{abstract}
We study a 6-dimensional brane world model with an abelian string
residing in the two extra dimensions. We study both static as well as 
inflating branes and find analytic solutions for the case of
trivial matter fields in the bulk. Next to singular space-times, 
we also find solutions which are regular including
cigar-like universes as well as solutions with periodic metric functions.
These latter solutions arise if in a singular space-time a static brane
is replaced by an inflating brane.
We determine the pattern of generic solutions for positive,
negative and zero bulk cosmological constant.    
\end{abstract}

\maketitle

\section{Introduction}
The idea that we live in more than the observed $4$ dimensions 
has been of huge interest ever since it was first suggested by Kaluza and Klein
in the 1920s \cite{kk}. They studied a five-dimensional gravitational
theory in a model with an extra {\it compact} dimension. The effective four dimensional
theory then contains $4$ dimensional gravity as well as the electromagnetic
fields and a scalar field. In a similar way, some (super)string theories 
contain solutions in which extra dimensions (required for the consistency
of those string theories) are all compact \cite{string} (superstring theories have
$6$ extra compact dimensions). The size of these compact dimensions is 
of the order of the
Planck length $=1.6 \cdot 10^{-33} cm$. An example is the
Calabi-Yau space in heterotic string theory whose properties
determine the low energy effective field theory.

Other models, which were discussed extensively in recent years are so-called
brane world scenarios \cite{ruba,dvali,anton,arkani,rs1,rs2,akama} which assume that the Standard model (SM) fields
are confined to a $3$-brane (a $3+1$ dimensional submanifold) which is
embedded in a higher dimensional space-time. Some of the extra dimensions
need now be {\it non-compact}.

Since gravity 
is a property of space-time itself, a model which describes appropriately the
well-tested Newton's law should localise gravity well enough to the $3$-brane.
This was achieved in \cite{rs2} by placing a $3$-brane into $5$ dimensions with the $5$th dimension
being infinite. For the localisation of gravity in this model, the brane tension has to be fine-tuned
to the negative bulk cosmological constant. 

Recently, the localisation of gravity on  different topological defects
has been discussed \cite{shapo}. 
This includes domain walls \cite{dfgk}, Nielsen-Olesen strings \cite{shapo1,shapo2}, 
and magnetic monopoles \cite{shapo3} in $5$, $6$ and 
$7$  space-time dimensions, respectively. It was found \cite{shapo,shapo1}
that gravity-localising (so-called ``warped'') solutions are possible
if certain relations between the defect's tensions hold.
While in the case of domain walls and strings, gravity can only be localised
when the bulk cosmological constant is negative, for magnetic monopoles
the gravity-localisation is possible for both signs of the cosmological constant.

Brane world models including higher dimensional topological defects have recently
also been discussed without bulk cosmological constant and from another point of view. 
7-dimensional brane worlds with a monopole (global and local) residing in
the three extra dimensions have been studied in \cite{cv}. 
The emphasis was put on the analysis of the localisation of the singularities in the
7-dimensional space-times. Inflating branes were studied, i.e. branes with a
positive 4-dimensional cosmological constant and it was shown that the singularity can
be removed from the space-time. 

In this paper we study the model of \cite{shapo1,shapo2,bh1,bh2} for inflating
branes. For completeness we also study the case of a positive bulk cosmological
constant and determine the pattern of solutions.

 Our paper is organised as follows : in section II, we present the 6 dimensional
Einstein-Abelian-Higgs (EAH) model,
we give the equations of motion and present special
analytic solutions for trivial matter fields.
In Section III, we present our numerical results.
We give our conclusions in Section IV.

\section{The model} 
We have the following $6$-dimensional action \cite{shapo}:
\begin{equation}
S_d=S_{gravity}+S_{brane}
\end{equation}
where the standard gravity action reads
\begin{equation}
\label{action_gravity}
S_{gravity}=-\int d^6 x 
\sqrt{-g} \frac{1}{16\pi G_{6}}\left(R+2\hat\Lambda_6\right) \ .
\end{equation}
$\hat\Lambda_6$ is the bulk cosmological constant, $G_6$ is the 
fundamental gravity scale with $G_6=1/M^4_{pl(6)}$ and
$g$ the determinant of the $6$-dimensional metric.

The action $S_{brane}$ for the 
Einstein--Abelian-Higgs (EAH) string is given in analogy to the 4-dimensional
case \cite{no,gstring} by:

\begin{equation}
S_{brane}=\int d^6 x \sqrt{-g_6} \left(-\frac{1}{4} F_{MN}F^{MN}
+\frac{1}{2}D_M\phi D^M\phi^*-\frac{\lambda}{4}(\phi^*\phi-v^2)^2  \right)
\end{equation}
with the covariant derivative $D_M=\nabla_M-ieA_M$ and the field strength
$F_{MN}=\partial_M
A_N-\partial_N A_M$ of the U(1) gauge potential $A_M$.
$v$ is the vacuum expectation value of the complex valued Higgs field $\phi$ and
$\lambda$ is the self-coupling constant of the Higgs field. 

\subsection{The Ansatz}
The Ansatz for the $6$-dimensional metric reads:
\begin{equation}
ds^2= M^2(\rho)\left[dt^2-\exp\left(2\sqrt{\frac{\Lambda_4}{3}}t\right)
\left(dr^2+r^2d\vartheta^2+r^2\sin^2\vartheta d\varphi^2\right)\right]
-d\rho^2- l^2(\rho)d\theta^2  \ ,
\end{equation}
where $\rho$ and $\theta\in [0:2\pi]$ are the coordinates associated
with the extra dimensions, while $r$, $\vartheta\in[0:\pi]$ and $\varphi\in[0:2\pi]$ are
the 4-dimensional coordinates. 
$\Lambda_4\geq 0$ is the 4-dimensional cosmological
constant. The 4-dimensional metric satisfies the 4-dimensional Einstein equations:
\begin{equation}
G_{\mu\nu}^{(4)}=8\pi G_{4}\Lambda_4 g_{\mu\nu}^{(4)}  \ \ , \ \ \mu,\nu=0,1,2,3 \ .
\end{equation}
The space-time has four Killing vectors:
$\frac{\partial}{\partial \theta}$, $\frac{\partial}{\partial \vartheta}$, 
$\frac{\partial}{\partial \varphi}$ and $\frac{\partial}{\partial t}$.
Here we have replaced the 4-dimensional flat Minkowski brane by an
inflating brane whose 4-dimensional metric is given by the
de Sitter metric. Note that we are using static coordinates here.
The 4-dimensional curvature scalar associated with this
inflating brane is then $R^{(4)}=4\Lambda_4$. The motivation to use
an inflating brane comes from the observationally confirmed
assumption that out universe has a small positive cosmological constant.

For the gauge and Higgs field, we have \cite{no}:
\begin{equation}
\phi(\rho, \theta)=v f(\rho) 
e^{i n\theta} \ , \ \ A_{\theta}(\rho,\theta)=\frac{1}{e}(n-P(\rho))
\end{equation}
where $n$ is the vorticity of the string, which throughout this paper we choose
$n=1$.
\subsection{Equations of Motion}
Introducing the following dimensionless coordinate $x$ and the dimensionless function $L$:
\begin{equation}
x=\sqrt{\lambda}v \rho \ , \ \ \ L(x)=\sqrt{\lambda}v l(\rho)
\end{equation}
the set of equations depends only on the following dimensionless coupling constants:
\begin{equation}
\alpha=\frac{e^2}{\lambda} \ , \ \ \ \gamma^2=8\pi G_6 v^2 \ , \ \ \ 
\Lambda=\frac{\hat\Lambda_6}{\lambda v^2} \ , \ \ \ 
\beta^2=\frac{\hat{\beta}^2}{\lambda v^4} \ , \ \ 
\kappa=\frac{8\pi G_4\Lambda_4}{\lambda v^2}
\end{equation}
The gravitational equations then read :
\begin{equation}
\label{eq1}
3 \frac{M^{''}}{M} + \frac{L^{''}}{L}
+ 3 \frac{L^{'}}{L} \frac{M^{'}}{M} 
+ 3 \frac{M^{'2}}{M^2}
+\Lambda-\frac{\kappa}{M^2}  =-\gamma^2 \left(\frac{(f^{'})^2}{2}+\frac{(1-f^2)^2}{4}+\frac{f^2 P^2}{2L^2}
+\frac{P'^2}{2\alpha L^2}   \right)
\end{equation}

\begin{equation}
\label{EQ2}
6 \frac{M^{' 2}}{M^2} + 
4 \frac{L^{'}}{L} \frac{M^{'}}{M} +\Lambda-2\frac{\kappa}{M^2}
= -\gamma^2 \left(-\frac{(f^{'})^2}{2}+\frac{(1-f^2)^2}{4}
+\frac{f^2 P^2}{2L^2} - \frac{P'^2}{2\alpha L^2} \right) \ ,
\end{equation}

\begin{equation}
\label{eq3}
6 \frac{M^{' 2}}{M^2}+
4 \frac{M^{''}}{M} +\Lambda -2\frac{\kappa}{M^2} =
-\gamma^2 \left(\frac{(f^{'})^2}{2}+\frac{(1-f^2)^2}{4}-\frac{f^2 P^2}{2L^2}
-\frac{P'^2}{2\alpha L^2} \right) \ .
\end{equation}
The Euler-Lagrange equations for the matter fields read:
\begin{equation}
\frac{(M^4 L f^{'})^{'}}{M^4L}+
(1-f^2)f-\frac{P^2}{L^2}f=0
\end{equation}
and
\begin{equation}
\label{peq}
\label{eqp}
\frac{L}{M^4}\left(\frac{M^4 P^{'}}{L}\right)^{'}-\alpha f^2 P=0 \ .
\end{equation}
The prime denotes the derivative with respect to $x$.

The equations (\ref{eq1})-(\ref{eq3}) can be combined to
obtain the following two differential equations for the two unknown metric functions:
\begin{equation}
\frac{(M^4 L^{'})^{'}}{M^4 L}+\frac{\Lambda}{2}
=\frac{\gamma^2}{2}\left(\frac{P'^2}{2\alpha L^2} - \frac{1}{4}(1-f^2)^2      \right)
\end{equation}
and
\begin{equation}
\frac{(LM^3 M^{'})^{'}}{M^4 L}+\frac{\Lambda}{2}-\frac{\kappa}{M^2}
=-\frac{\gamma^2}{2}\left(\frac{2 P^2 f^2}{L^2} + \frac{1}{4}(1-f^2)^2
+\frac{1
}{2} \frac{P'^2}{\alpha L^2}            \right)  \ .
\end{equation}

\subsection{4-dimensional effective action}
In the following we will discuss the 4-dimensional effective action
arising in the limit $f(x)\equiv 1$, $P(x)\equiv 0$.
The case where the matter fields are equal to their vacuum values is
of interest to understand the asymptotic behaviour of the system in the generic
case. Many publications have dealt with this, e.g. \cite{olo}, since in contrast
to the generic case analytic results are available.

Note that from (\ref{EQ2}) and (\ref{eq3}) it follows
that $L(x)\propto M'(x)$ unless $M(x)$ is constant.

Due to the Ansatz we use for the metric, a dimensional reduction
of the gravity 
action (\ref{action_gravity}) can be performed. For this purpose notice that
we can write the 6-dimensional Ricci scalar $R$ according to:
\begin{equation}
R=\frac{R^{(4)}}{M^2} - 8\frac{M''}{M} - 12\frac{M'^2}{M^2} - 8\frac{L' M'}{LM}
-2 \frac{L''}{L}
\end{equation} 
where $R^{(4)}$ is the 4-dimensional
Ricci scalar associated with the metric on the brane.
Using the fact that $L(x)=\tilde{c}M'(x)$, $\tilde{c}$ constant, we obtain the following
4-dimensional effective action:

\begin{equation}
\label{action_gravity_effective}
S_{gravity}=-\int d^4 x \sqrt{-g^{(4)}} 
\frac{1}{16\pi G_{eff}}\left(R+\Lambda^{(4)}_{eff}\right) \ ,
\end{equation}
where 
\begin{equation}
\Lambda^{(4)}_{eff}=\left.\frac{2\tilde{c}\pi G_{eff}}{G_6}\left(\frac{2}{5} \hat{\Lambda}_6 M^5 - 
2 M'' M^4 -4 M^3 M'^2\right)
\right\vert_{x=x_1}^{x=x_2}
\end{equation}
and  at first order, taking into account the fact that the
matter fields are localized and will not influence the asymptotic form,
we can define an effective coupling constant according to
\begin{equation}
\label{plmas}
M_{pl}^2 = \frac{1}{G_{eff}}=\left.\frac{2\tilde{c}\pi}{3G_6} M^5\right
\vert_{x=x_1}^{x=x_2} \ ,
\end{equation}
where $x\in [x_1:x_2]$. $x_2-x_1$ is then the radius of the 2-dimensional
transverse space. The value of the parameter $\tilde c$ can be
approximated by its value in absence of matter fields which can be
evaluated exactly to be $\tilde c = 4/\vert 2\kappa - \Lambda \vert$.

\subsection{Special solutions}
Explicit solutions can be constructed for $f(x)\equiv 1$ and $P(x)\equiv 0$.
These are by themselves of interest, of course, but are also interesting for the
generic solutions since we would expect that the metric fields take the form
presented below far away from the core of the string.
A similar analysis has been done for global defects in \cite{olo}.

\subsubsection{Static branes $\kappa=0$}
For $\Lambda=0$ the system admits two different types of solutions \cite{bh1,bh2}:
\begin{equation}
\label{sol1}
             M_S = 1 \ \ , \ \ L_S = x - x_0
\end{equation}
and
\begin{equation}
\label{sol2}
             M_M = \tilde{C_1}(x-\tilde{x_0})^{2/5} \ \ , 
\ \ L_M = \tilde{C_2}(x-\tilde{x_0})^{-3/5}
\end{equation}
where $\tilde{x_0}$, $\tilde{C_1}$, $\tilde{C_2}$ are arbitrary constants.
By analogy to the 4-dimensional case, we refer to these solutions
as to the ``string'' (S) and ``Melvin'' (M) branches, respectively. Note that the only difference
to the 4-dimensional case are the powers in (\ref{sol2}).

For $\Lambda > 0$ the explicit solutions read:

\begin{equation}
\label{vacuumtri}
             M = C_1 \sin\left(\lambda (x- x_0)\right)^{2/5} \ \ , \ \
             L = C_2 \frac{\cos \left(\lambda (x-x_0)\right)}
             {\sin \left(\lambda (x-x_0)\right)^{3/5}}
\end{equation}
where $C_1$, $C_2$, $x_0$ are arbitrary 
parameters and $\lambda^2 \equiv  5 \Lambda/8$.
These solutions are periodic in the metric functions. In the 4-dimensional
analogue, i.e. Nielsen-Olesen strings in de Sitter space, periodic
solutions also appear for trivial matter fields \cite{bbh,linet}.

It is easy to see that the string and Melvin solutions (\ref{sol1}), (\ref{sol2}) can be obtained
from these solutions
for special choices of the free parameters and specific limits
(for $\Lambda \to 0$) of the trigonometric solution.

The solutions for $\Lambda < 0$ are given by
\begin{equation}
             M = A_{\pm} \exp (\pm \sigma x) \ \ , \ \ 
             L = B_{\pm} \exp (\pm \sigma x) \ \ , \ \  
             \sigma^2 = -\frac{\Lambda}{10}
\end{equation}
where $A_{\pm}$, $B_{\pm}$  are arbitrary constants.

\subsubsection{Inflating branes $\kappa\neq0$}
In the case $\kappa >0$ the solution for $M$, $L$
can be written in terms of quadratures as follows:
\begin{equation}
\label{quadrature}
       x - \hat{x}_0 =  \int dM \sqrt{\frac {M^3}{{\frac{\kappa}{3}M^3
       - \frac{\Lambda}{10}M^5+C}}}
  \ \ , \ \ L = \tilde{L}_0 \frac{dM}{dx}  = 
  \tilde{L}_0 \sqrt{ \frac{  \frac{\kappa}{3}M^3
              - \frac{\Lambda}{10} M^5
               + C} {M^3}  
               }
\end{equation}
where $\hat{x}_0$, $C$ are integration constants and $\tilde{L}_0$ is arbitrary.
The explicit form of $M$ is involved. It   depends
on elliptic functions  if $\Lambda = 0$, in the case $\kappa=0$,
the integration can be done by an elementary change of variable and
leads to the expressions above.
In the particular case $\Lambda=0$ and with $C=0$ we find
\begin{equation}
\label{kappasol}
     M = \sqrt{\frac{\kappa}{3}}(x-\hat{x}_0) \ \ , \ \ 
L = L_0 \equiv \tilde{L}_0 \sqrt{\frac{\kappa}{3}}= {\rm constant}  \ .
\end{equation}
This latter solution corresponds to a cigar-type solution
(in the extra dimensions) and we find
that for $C\neq 0$ the solutions are also of this type, however cannot
be given in an explicit form.
``Cigar-type'' refers to the fact that $g_{\theta\theta}=const.$.
Consequently, the circumference of a circle in the two extra dimensions
becomes independent of the bulk radius $\rho$.

An analogue solution
in $7$ dimensions with monopoles residing in the three extra dimensions
has been found previously in \cite{cv}.
In the case $\Lambda > 0$, $\kappa > 0$ and $C=0$ 
we find
\begin{equation}
\label{trigonometric}
      M = \sqrt{\frac{10 \kappa}{3 \Lambda}} 
      \sin\left(\sqrt{\frac{\Lambda}{10}}(x-\hat{x}_0)\right)  \ \ , \ \
 L = L_0 
      \cos\left(\sqrt{\frac{\Lambda}{10}}(x-\hat{x}_0)\right)  \ .
\end{equation}
Again, the metric functions are periodic. The periodicity of string-like solutions
in de Sitter space seems to be a generic feature - independent of the number of space-time
dimensions  or of the type of brane present. 

\subsection{Boundary conditions}
We require
regularity at the origin $x=0$ which leads to the boundary conditions:
\begin{equation}
\label{bcx0}
f(0)=0 \ , \ \ \ P(0)=1 \ , \ \ \ M(0)=1 \ , \ \ \ M^{'}|_{x=0}=0 \ , \ \ \
L(0)=0 \ , \ \ \  L^{'}|_{x=0}=1 \ .
\end{equation}
Along with \cite{shapo1,shapo3,cv}, we assume the matter fields
to approach their usual asymptotic value far from the string core
(i.e. for $r  >>  \sqrt{\lambda} v$)~:
\begin{equation}
f(x \to \infty)=1 \ , \ \ \ P(x \to \infty)=0 \ .
\end{equation}

\subsection{Asymptotic behaviour}
Close to the origin $x=0$, the functions have the following behaviour:
\begin{eqnarray}
f(x<<1)&\simeq & f_0x+f_0\left(\frac{1}{12}f_0^2 \gamma +\frac{\gamma}{6\alpha} 
p_0+ \frac{1}{4}p_0+\frac{1}{48} \gamma + \frac{1}{12} \Lambda
-\frac{1}{8} - \frac{1}{16} \kappa\right)x^3 \ , \\
P(x<<1)&\simeq & 1+p_0 x^2 \ , \\
M(x<<1)&\simeq & 1+ \left(\frac{1}{4} p_0^2 \frac{\gamma}{\alpha}-\frac{1}{32}
\gamma -\frac{1}{8}\Lambda+ \frac{3}{16}\kappa  \right) x^2 \ , \\
L(x<<1)&\simeq &  x+ \left(-\frac{1}{6} f_0^2\gamma -\frac{5}{6} p_0^2 \frac{\gamma}{\alpha}
+\frac{1}{48} \gamma +\frac{1}{12}\Lambda -\frac{1}{4}\kappa   \right) x^3
\end{eqnarray}
For $\kappa=0$ the behaviour found in \cite{shapo1} is recovered.

For $x >>1$, the matter functions tend to their asymptotic values 
$P(x >>1)\to 0$,
$f(x >>1)\to 1$. The metric functions thus tend asymptotically
to the special solutions described above.

\section{Numerical results}
Following the investigations in \cite{cv}, we here mainly aim at a classification
of the generic solutions available in the system.
\subsection{Static branes ($\kappa=0$)}
We solve the system of ordinary differential equations
subject to the above boundary conditions numerically.
The system depends on three independent coupling constants $\alpha$,
$\gamma$, $\Lambda$. We here fix $\alpha=2$, corresponding to the
self dual case in flat space, and we will in the following describe
the pattern of solutions in the
$\gamma$-$\Lambda$ plane. Results for the
4-dimensional gravitating string \cite{gstring} make us believe 
that the pattern
of solutions for $\alpha\neq 2$ is qualitatively similar. For vorticity $n >1$,
the situation might change (see \cite{gstring}), however, 
we don't discuss this case here.
As will become evident, the presence of two cosmological constants
leads to a rather complicated pattern of solutions. 
The different types of solutions available are summarized is the Summary
section. Their interconnection is illustrated in Fig.\ref{fig11}.

\subsubsection{Zero or 
negative bulk cosmological constant}

\begin{figure}
\epsfysize=10cm
\epsffile{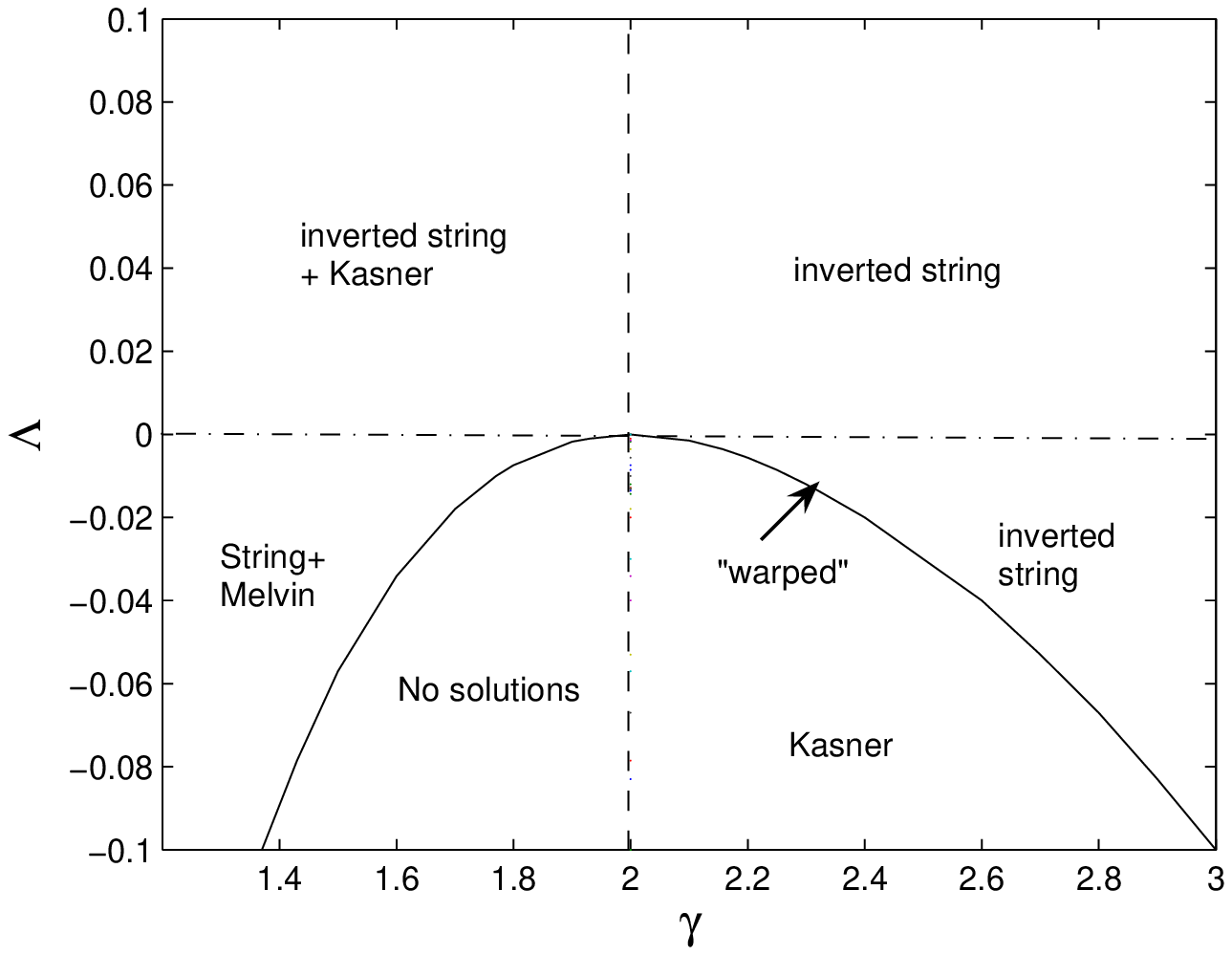}
\caption{\label{fig1}
The pattern of brane world solutions 
in the $\Lambda$-$\gamma$ plane for $\alpha=2$.
}
\end{figure}

The case $\Lambda=0$ was studied in detail in \cite{bh1,bh2}. We review the
main results here to fit it into the overall pattern of solutions. The
pattern of solution in the $\gamma$-$\Lambda$-plane is given in Fig.\ref{fig1}.

For $\Lambda=0$ and $\gamma < 2$ two branches of solutions exist,
with an asymptotic behaviour
of the metric functions given by (\ref{sol1}) and  (\ref{sol2}).
Referring to their counterparts in a
 four-dimensional space-time \cite{gstring}
we denote these two families of solutions as
the ``string'' and ``Melvin'' branches, respectively. The terminology
used e.g. in \cite{gstring} will be used throughout the rest of the paper.

\begin{figure}
\epsfysize=22cm
\epsffile{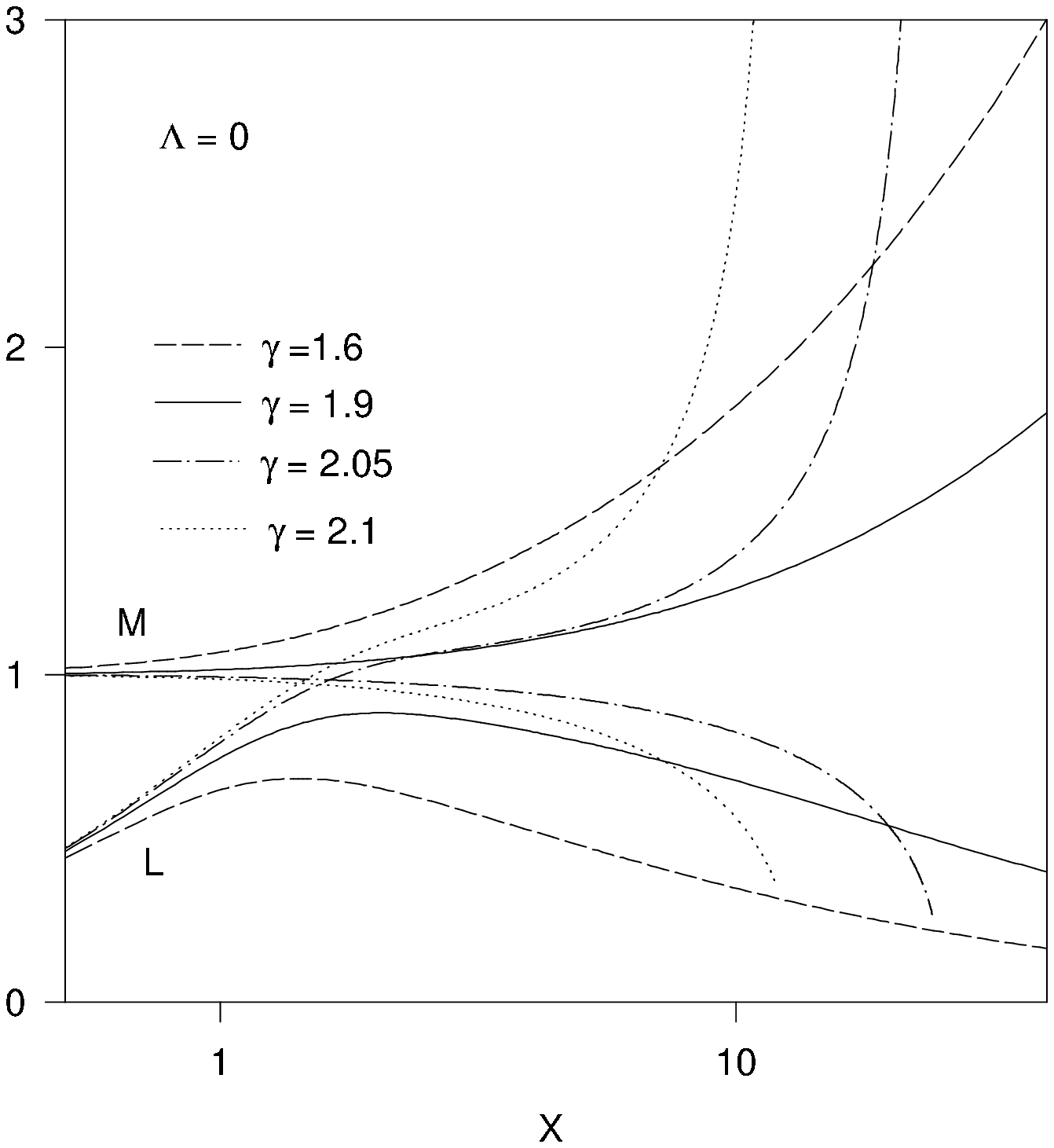}
\caption{\label{fig2}
The pattern of the transision from the Melvin solution ($\gamma < 2$)
to the Kasner solution ($\gamma > 2$) for $\Lambda = 0$ if given for the metric functions $M$ and $L$.
}
\end{figure}

Specifically for $\alpha=2$, we have $M(x)\equiv 1$ and
$L(x>>1) \sim ax+b$, $a>0$.
At $\gamma = 2$ the two solutions coincide and $L(x >> 1)=1$.
When the parameter $\gamma$ is increased to values larger than $2$
the string and Melvin solutions get progressively deformed into
closed solutions with zeros of the metric functions.
For $\gamma > 2$ indeed, the
the string branch smoothly evolves into the so called
inverted string branch (again using the terminology of \cite{gstring}).
The inverted string solutions are characterized by the fact that the
slope of the function $L(x)$ is constant and negative, $L(x)$ therefore
crosses zero at some finite value of $x$, say $x=x_{IS}$.
On the other hand, the Melvin branch evolves
into the Kasner branch, a configuration for which the function $M(x)$
develops a zero at some finite value of $x=x_K$ while $L(x)$ becomes
infinite for $x\to x_K$. More precisely,
 for $0 << x < x_K$  these solutions have the behaviour
$M \propto (x_K-x)^{2/5}$, $L \propto (x_K-x)^{-3/5}$.
The transition between Melvin solutions
(for $\gamma < 2$) and Kasner solutions
(for $\gamma > 2$) is illustrated in Fig. \ref{fig2}.

For negative $\Lambda$ and $\gamma < 2$, the string and Melvin solutions
are still present and merge into a single
solution at some critical value of $\Lambda$.  For $\Lambda <0$ and
$\gamma > 2$, we have the zero $x_{IS}$ of the inverted
string solution increasing with decreasing $\Lambda$. It reaches infinity for
the exponentially decreasing solution, the so-called ``warped'' solution
that localises gravity on the brane \cite{bh1,bh2}. 
If the cosmological constant is further
decreased the solution becomes of Kasner-type.

\subsubsection{Positive bulk cosmological constant ($\Lambda >  0$)}

Up to now static branes have only been discussed in an asymptotically
flat or Anti-de Sitter bulk. Here, we also consider the case of static branes
in a de Sitter bulk. 
First, we examine the evolution of the string and Melvin solutions for
$\Lambda > 0$. For all solution constructed with $\Lambda > 0$, we were
able to recover the behaviour (\ref{vacuumtri}) asymptotically.

\begin{figure}
\epsfysize=22cm
\epsffile{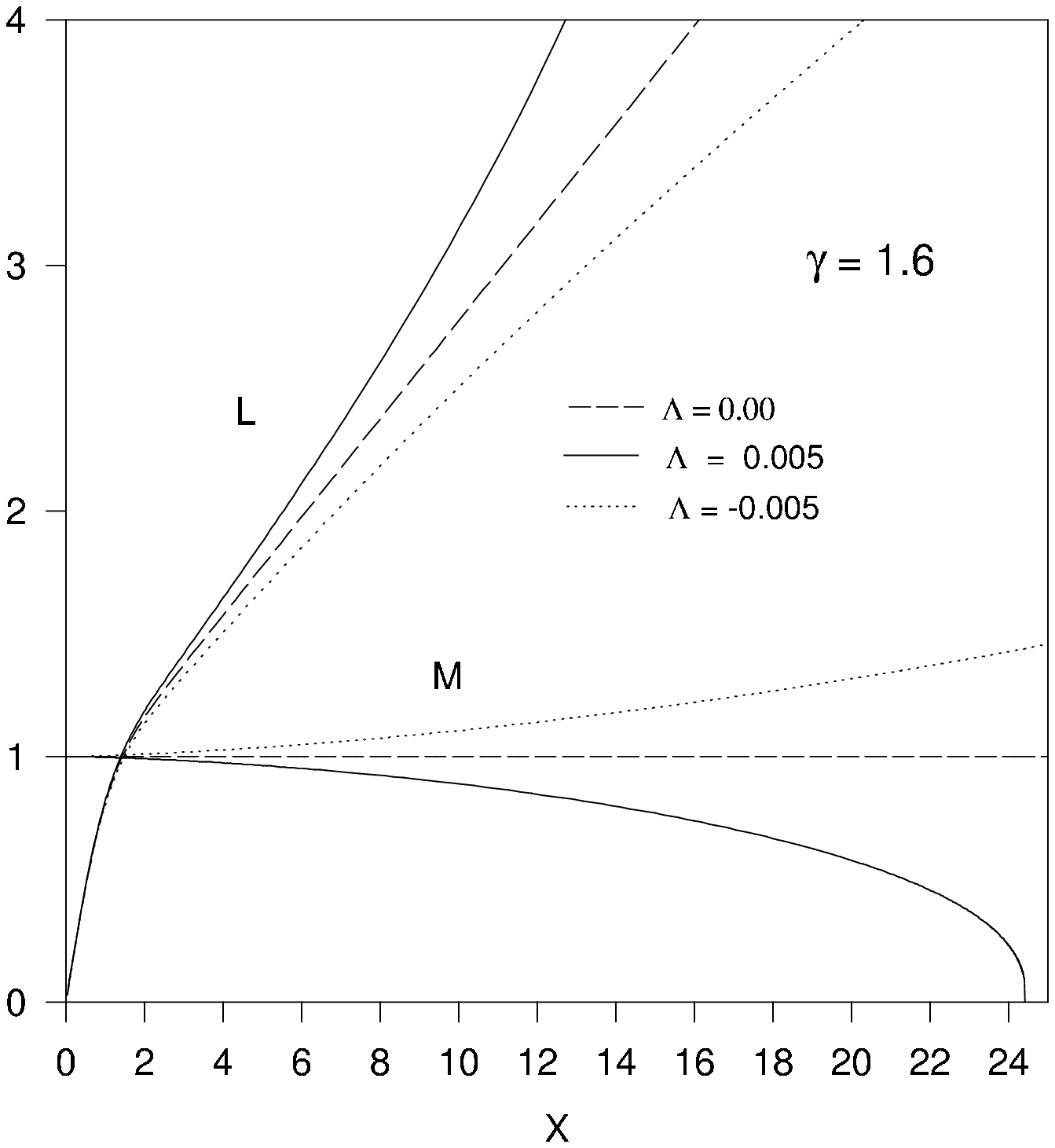}
\vskip -3cm
\caption{\label{fig3}
The profiles of the metric functions $M$, $L$  of the
string solution (for $\Lambda=0$, $\Lambda=-0.005$), respectively of
the Kasner solution (for $\Lambda=0.005$) for $\gamma=1.6$. 
}
\end{figure}

\begin{figure}
\epsfysize=22cm
\epsffile{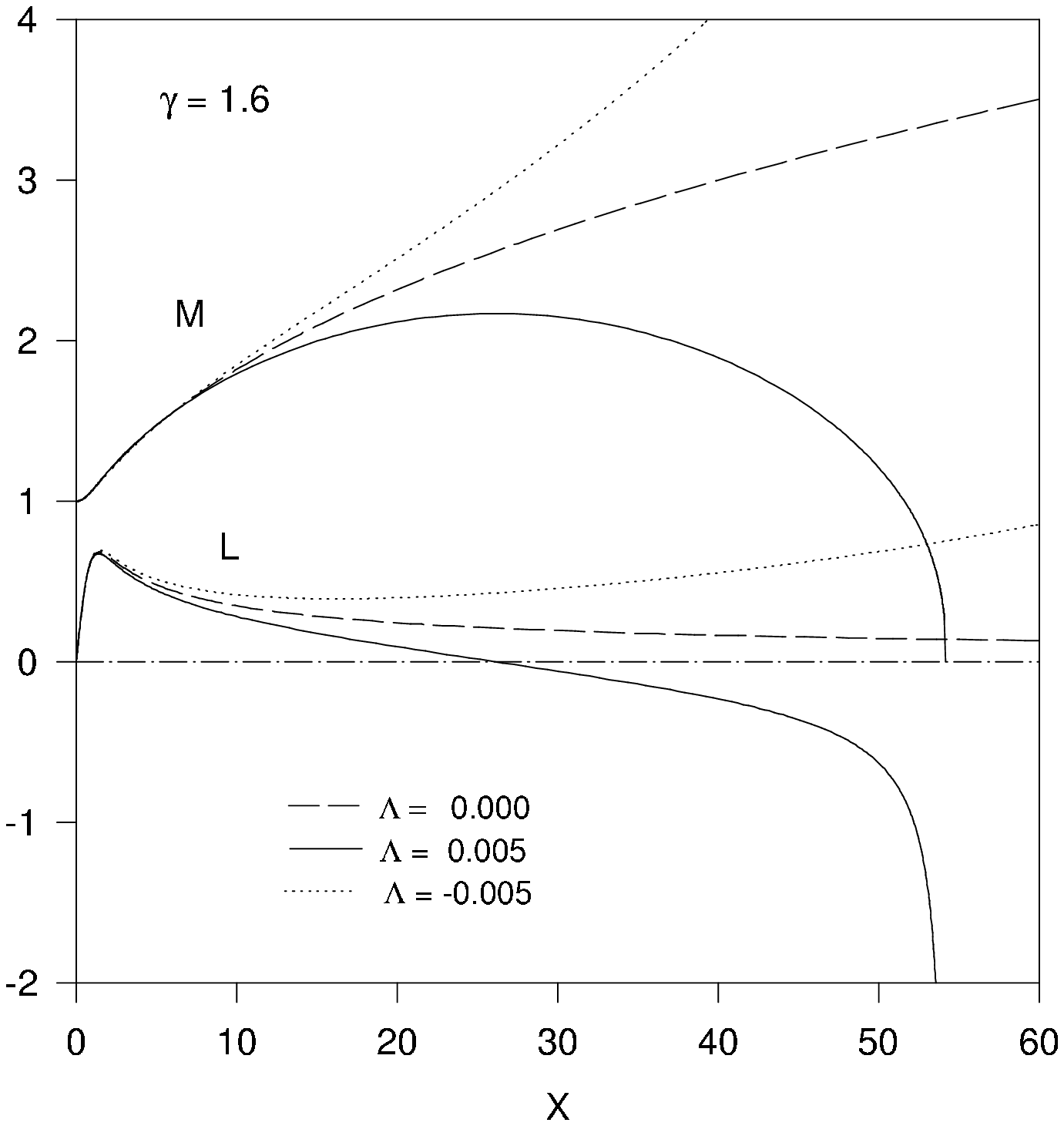}
\vskip -3cm
\caption{\label{fig4}
The profiles of the metric functions $M$, $L$  of the
Melvin solution (for $\Lambda=0$, $\Lambda=-0.005$), respectively
of the inverted string solution (for $\Lambda=0.005$) $\gamma=1.6$.
}
\end{figure}

This evolution is illustrated in Fig.\ref{fig3} and  Fig.\ref{fig4}
for $\gamma = 1.6$ respectively for the string and Melvin solutions.
Here, we show the metric functions $M$, $L$
for  $\Lambda = 0$ and $\Lambda = \pm 0.005$.
For $\Lambda > 0$ we find a solution with the  metric function $M(x)$
possessing a zero at some finite $x$, say $x=x_1$.
At the same
time $L(x\to x_1)\to + \infty$.  These solutions tend to the
string solutions in the limit $\Lambda \to 0$.
Following the convention used in the $\Lambda=0$-case,
we refer to these solutions
as of ``Kasner''-type. 

The second type of solutions that we find has metric functions
behaving for $0< x_2 < x_3$ as
\begin{equation}
              L(x_2)= 0 \ \ , \ \ M'(x_2)=0 \ \ , \ \
              M(x_3)= 0 \ \ , \ \ \lim_{x \to x_3} L(x) = - \infty  \ .
\end{equation}
These tend to the Melvin solutions in the limit $\Lambda\to 0$ and
we will refer to them as of ``inverted string''-type.

\begin{figure}
\epsfysize=22cm
\epsffile{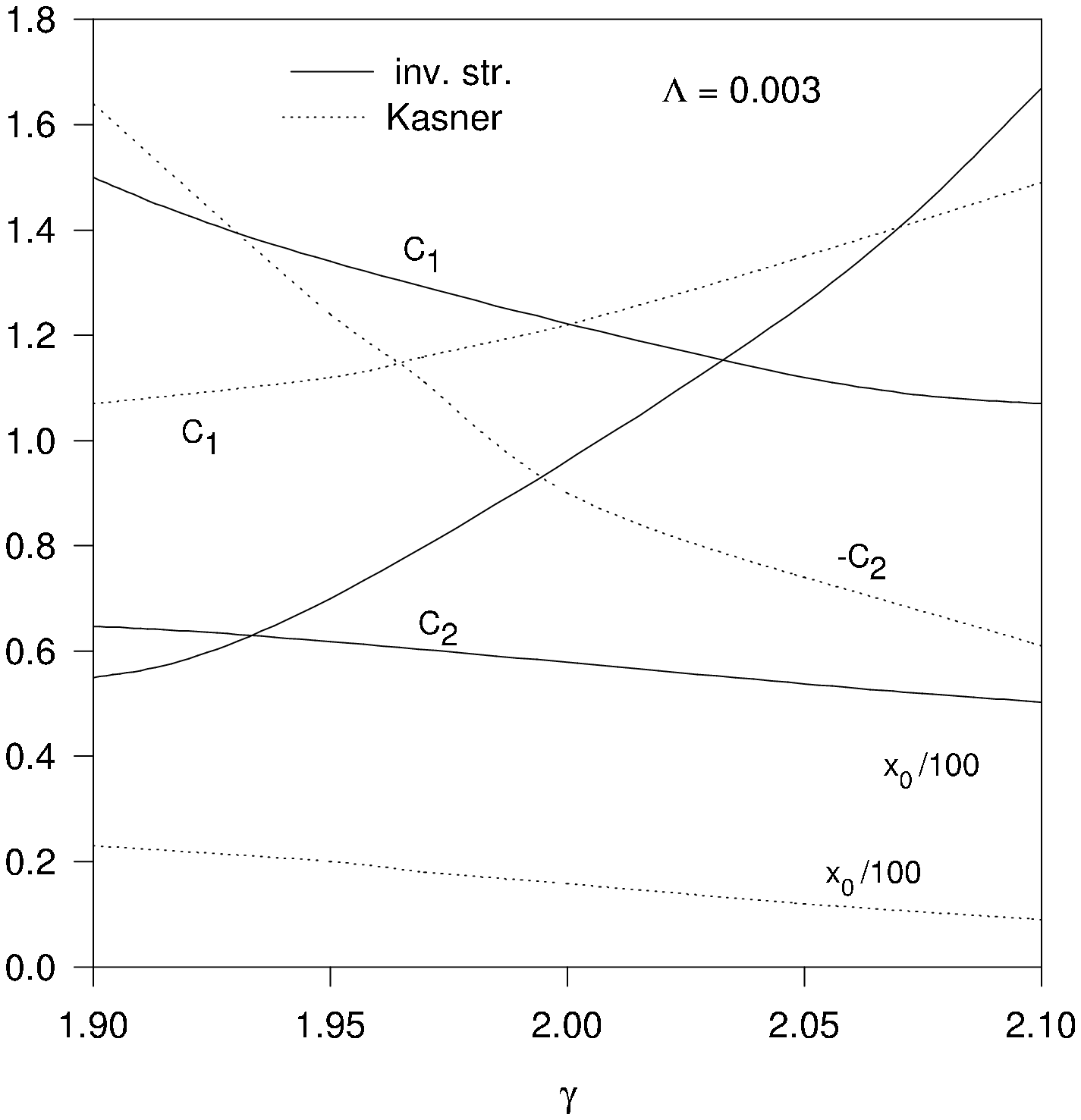}
\vskip -3cm
\caption{\label{fig5}
The values of the parameters $C_1,C_2,x_0$ defined in the text
for the inverted string and Kasner-like solutions as functions
of $\gamma$ for $\Lambda = 0.003$
}
\end{figure}

For fixed $\Lambda > 0$ and
increasing $\gamma$ we find that both Kasner and inverted string solutions
exist for all values of $\gamma$.
This is demonstrated in Fig.\ref{fig5} where the values
of the parameters $C_1$, $C_2$, $x_0$ 
(defined in Eq.(\ref{vacuumtri}))
are plotted as functions of $\gamma$.

 \subsection{Inflating branes $\kappa > 0$}
Here, we discuss inflating branes ($\kappa > 0$).
Again, we fix $\alpha=2.0$. Since observational evidence points
to a positive cosmological constant in our universe (which is represented
by the inflating 3-brane) we restrict to $\kappa > 0$ here.

The pattern of solutions can largely be characterized
by the integration constant $C$ appearing in (\ref{quadrature}).

\subsubsection{Zero bulk cosmological constant}

\begin{figure}
\epsfysize=22cm
\epsffile{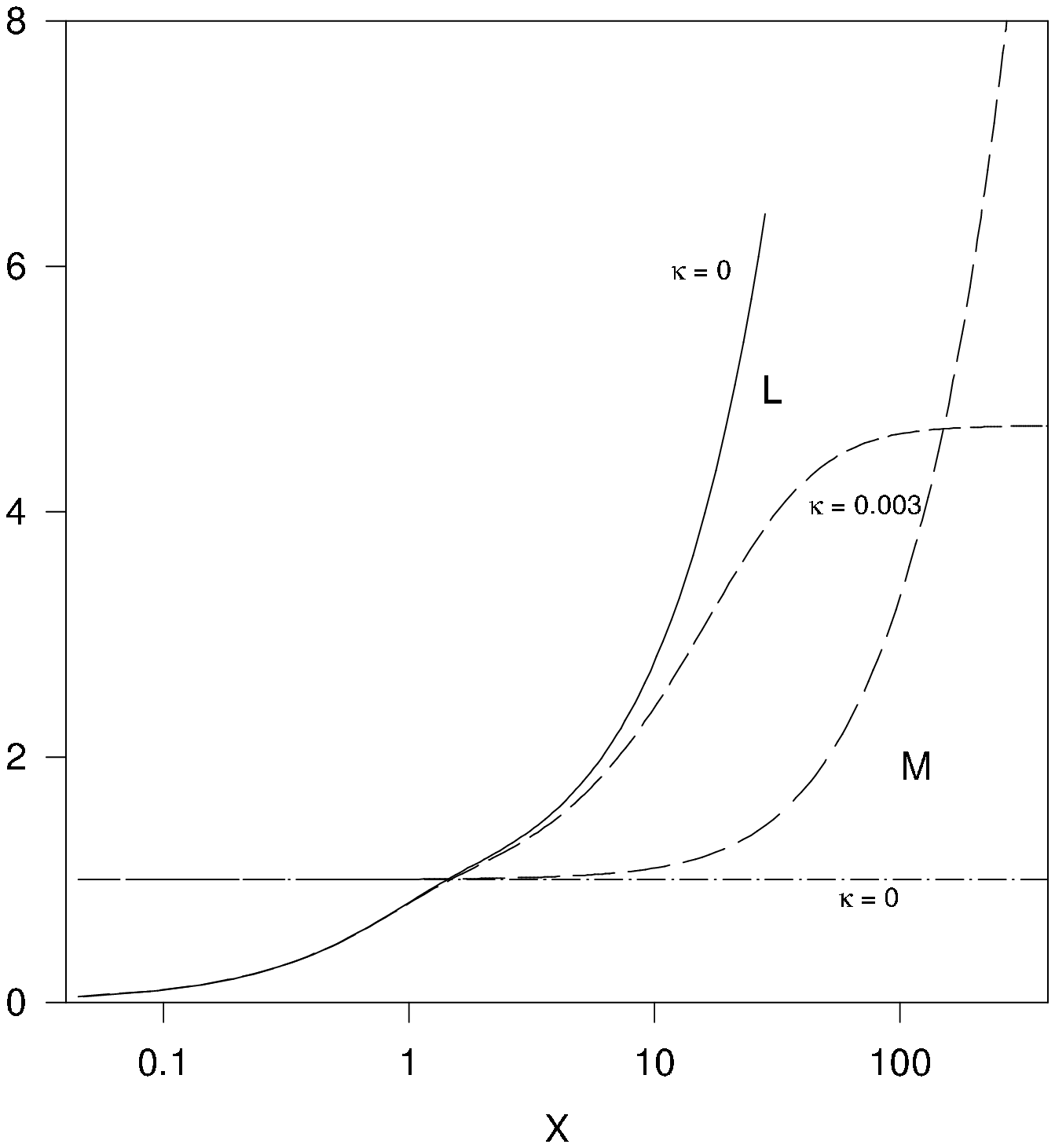}
\vskip -3cm
\caption{\label{fig6}
The profiles of the metric functions $M$, $L$ for $\Lambda= 0$
and for $\kappa=0$ and $\kappa = 0.003$, respectively.
}
\end{figure}

First, we have constructed solutions corresponding to deformations of the 
string solutions residing in the extra dimensions. We present
the profiles in Fig. \ref{fig6} for $\kappa = 0.003$ and for comparison
for $\kappa=0$.
Obviously, the presence of an inflating brane ($\kappa>0$) 
changes the asymptotic behaviour of $M$, $L$ drastically in comparison
to a Minkowski brane.
$M(x)$ now behaves lineary far from the core of the string, while 
$L(x)$ tends to a constant. The solutions 
approach asymptotically (\ref{quadrature}). The space-time is thus cigar-like. 

This can be explained as follows~:
In the case $\alpha = 2$ the equations are self dual, in particular
the equation determining the function $M$ on the string branch is
$M M'' + (3/2) (M')^2 - \kappa/2 = 0$ (the combination of the 
energy momentum tensor vanishes identically for $\alpha=2$ and for 
the string like solution).
The value of $C$ compatible with the boundary condition 
turns out to be $C = - \kappa/4$,
since we are interested is small values of $\kappa$, the integral
(\ref{quadrature}) can be reasonably approximated by (\ref{kappasol}),
in complete agreement with our numerical results.

\begin{figure}
\epsfysize=12cm
\epsffile{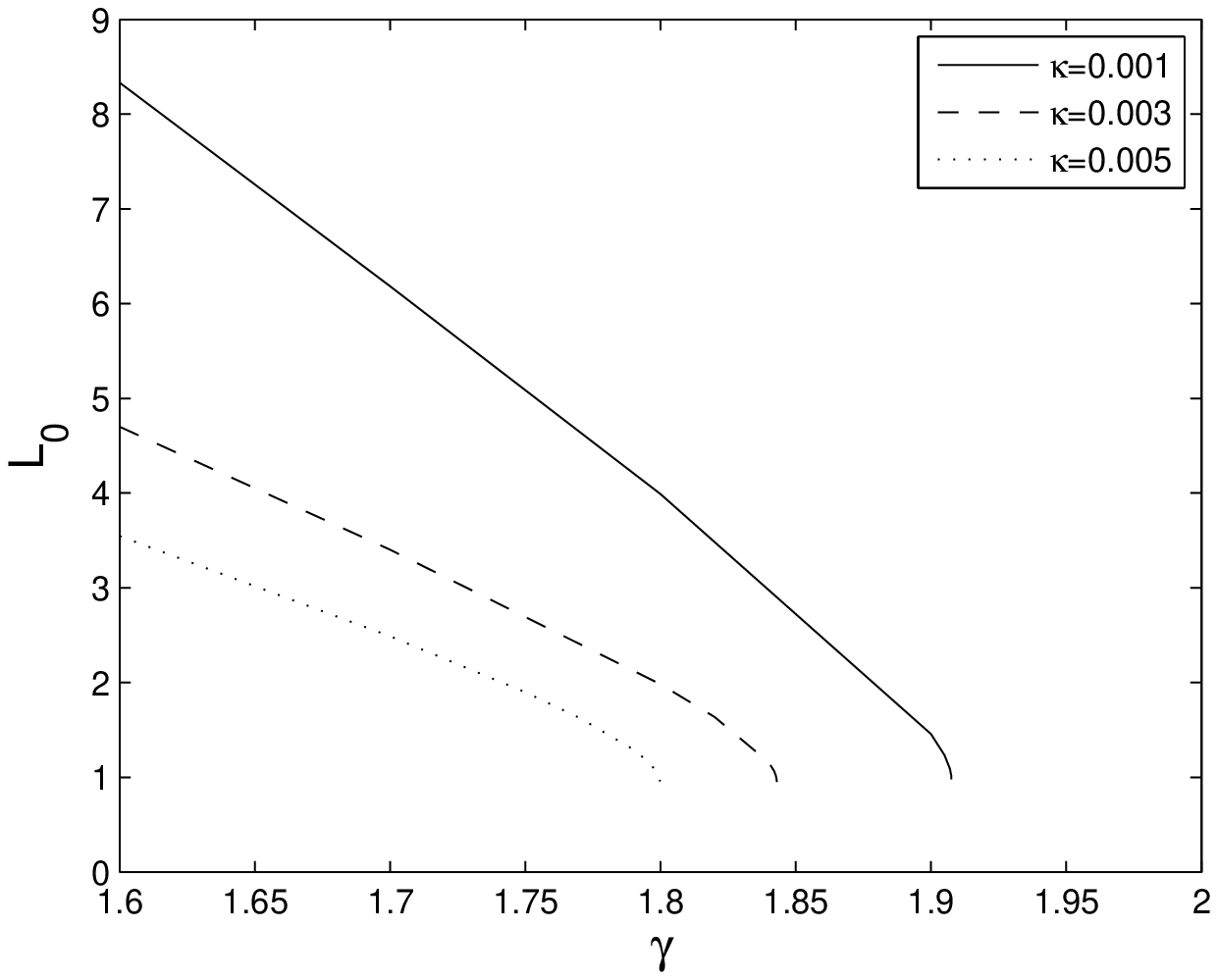}
\vskip 0cm
\caption{\label{fig7}
The parameter $L_0$ of the cigar-like solution
as function of $\gamma$ for different values of $\kappa$ and $\Lambda=0$. }
\end{figure}

\begin{figure}
\epsfysize=12cm
\epsffile{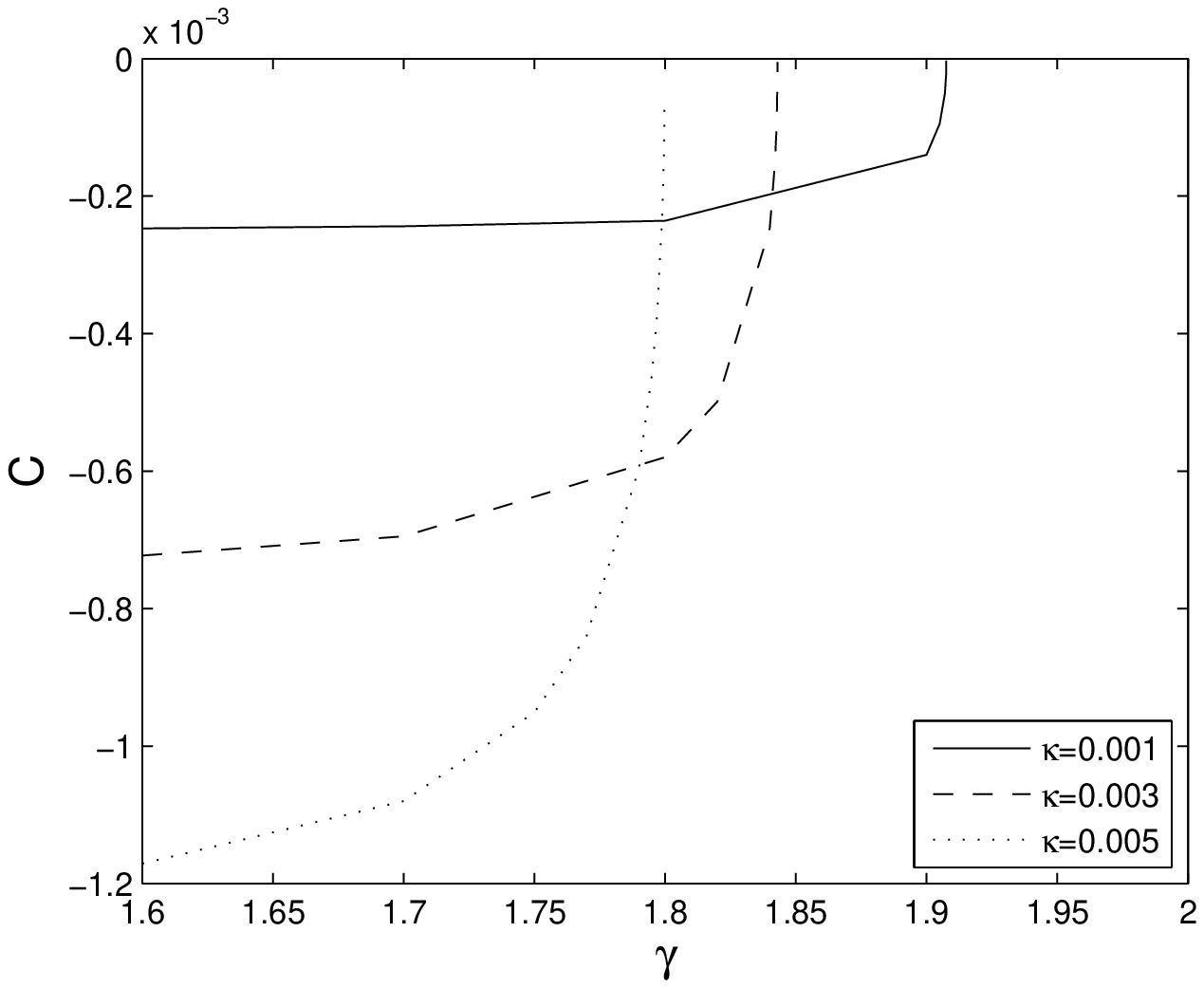}
\vskip 0cm
\caption{\label{fig8}
The parameter $C$ of the cigar-like solution
as function of $\gamma$ for different values of $\kappa$ 
and $\Lambda=0$.}
\end{figure}

The parameters $L_0$ and $C$ appearing in (\ref{quadrature}) can be determined
numerically. It turns out that for the cigar-like solutions we always 
have $C < 0$, while $L_0$ is positive. For a fixed value of $\kappa$
and varying $\gamma$, we find that $C\to 0$ for a critical, $\kappa$-dependent 
value of $\gamma$, $\gamma_{cr}(\kappa)$. At the same time $L_0 \to 0$ for
$\gamma \to \gamma_{cr}$. This is shown for three different values 
of $\kappa$ in Fig.\ref{fig7} and Fig.\ref{fig8}. 
In the limit $\gamma \to \gamma_{cr}$, the diameter
of the cigar tends to zero and the cigar-like solutions cease to exist.

We have also studied solutions corresponding to 
deformations of the Melvin solutions.
It turns out that the $\kappa = 0$ Melvin solution is smoothly
deformed for $\kappa > 0$.  $C$ is always positive
and tends exponentially to zero as function of $\gamma$. 
For $\gamma > 2$ we find that the solutions have a zero of the 
metric function $L$ at
some $\gamma$- and $\kappa$-dependent value of $x$. The solutions are thus
of inverted string-type.

\begin{figure}
\epsfysize=12cm
\epsffile{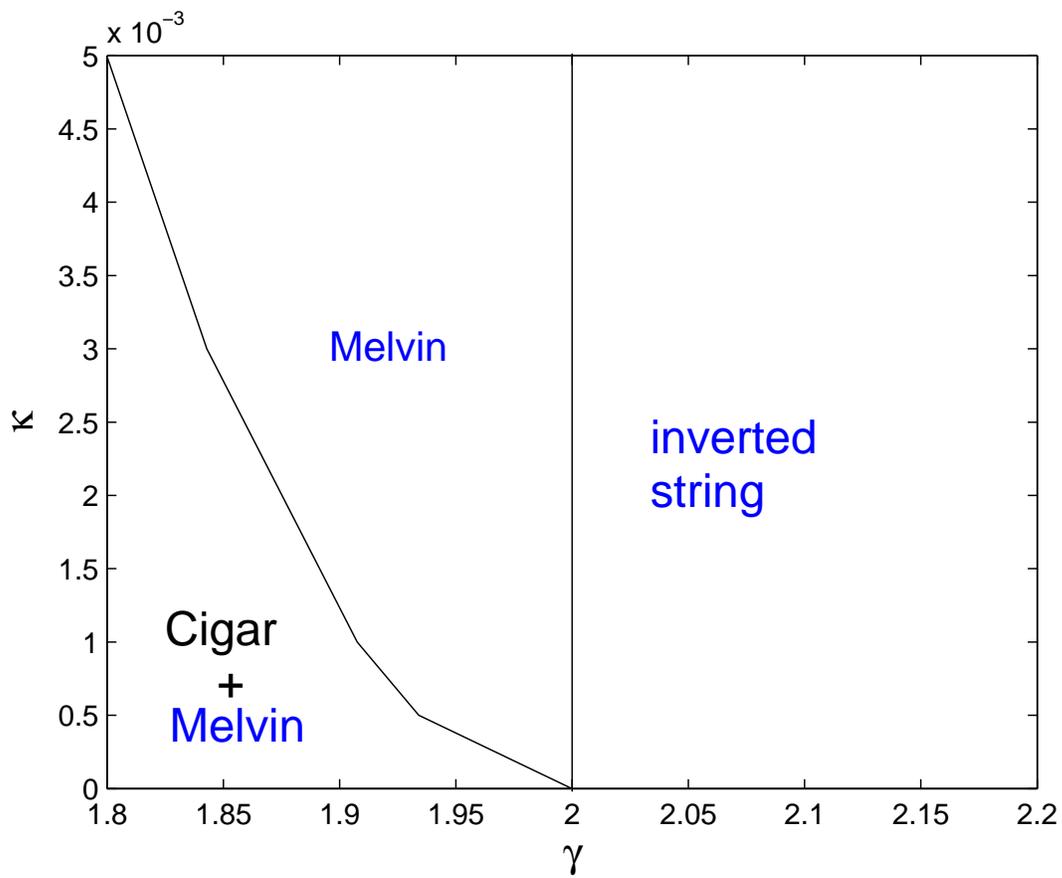}
\vskip 0cm
\caption{\label{fig9}
The pattern of inflating brane solutions in the $\gamma$-$\kappa$-plane.}
\end{figure}

We present the pattern of solutions in the $\gamma$-$\kappa$-plane
in Fig.\ref{fig9}.

\subsubsection{Positive/negative bulk cosmological constant}
We have limited our analysis here to 
$\gamma \leq 1.8$. For $\gamma > 1.8$, the numerical analysis
becomes very unreliable, that is why we don't report our results here.

Fixing $\gamma$, e.g. to $\gamma=1.8$, 
the inverted string solutions (available for $\Lambda > 0$, $\kappa=0$) 
gets smoothly deformed for $\kappa > 0$. The constant $C$ is positive
for all solutions. 
For fixed $\kappa$ and $\Lambda \to 0$ the Melvin solution is approached.

\begin{figure}
\epsfysize=22cm
\epsffile{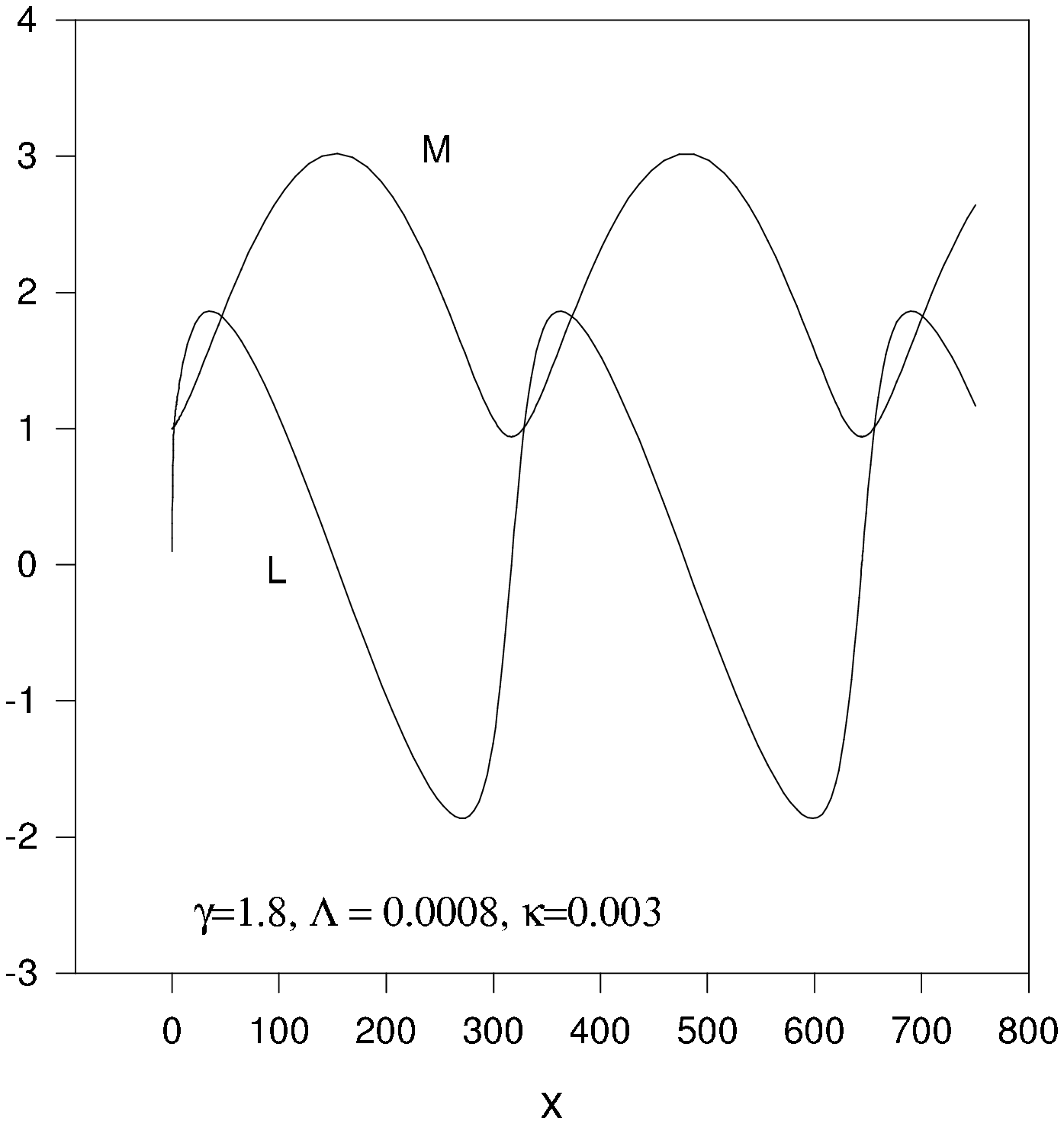}
\vskip -3cm
\caption{\label{fig10}
The profiles of the metric functions $M$, $L$ for $\Lambda= 0.0008$
and for $\kappa=0.003$ and $\gamma = 1.8$.
}
\end{figure}

\begin{figure}
\epsfysize=10cm
\epsffile{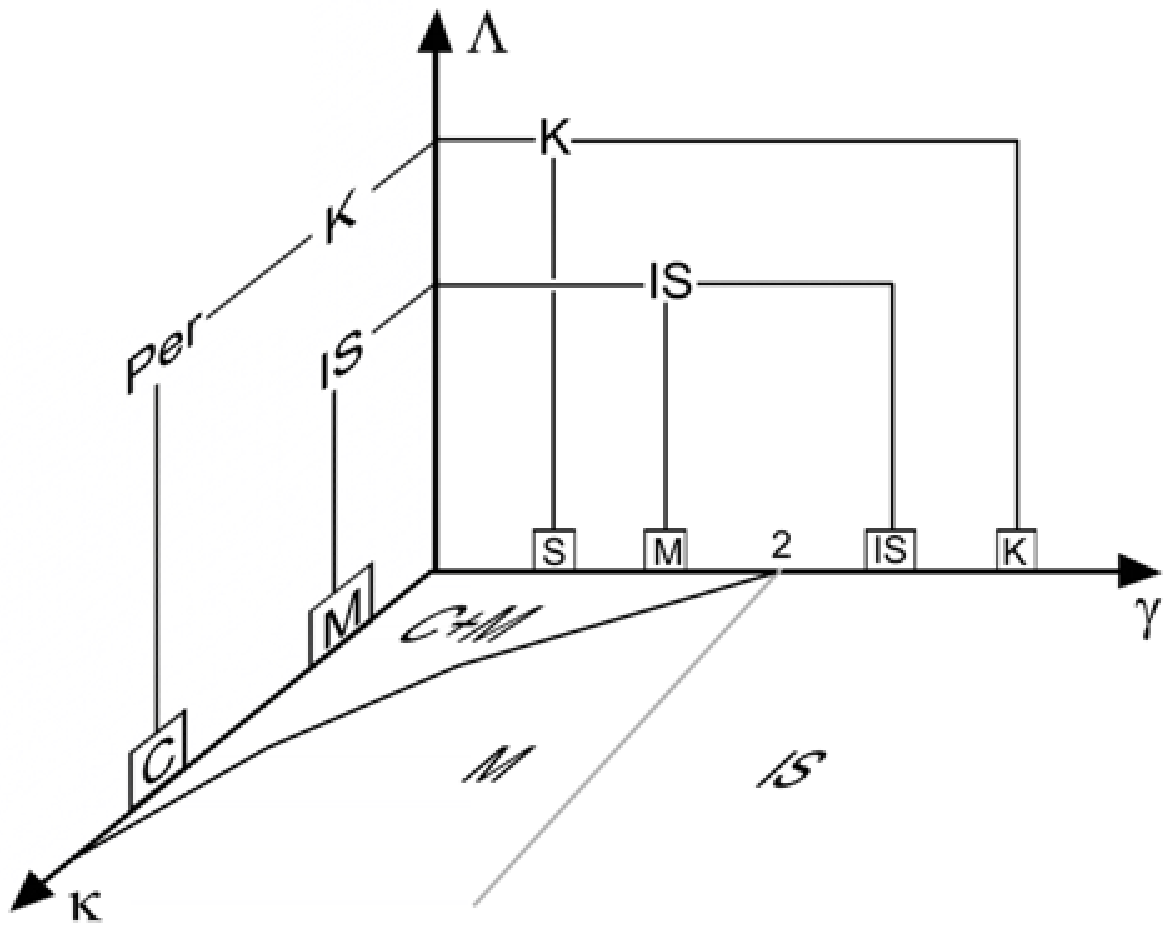}
\vskip 3cm
\caption{\label{fig11}
The pattern of brane solutions in the $\kappa$-$\gamma$-$\Lambda$-domain.
The origin of the coordinate system here corresponds to $\kappa=0$, $\gamma=1.8$, $\Lambda=0$.
M, S, IS, K, Per, C denote Melvin, string, inverted string, Kasner, periodic
and cigar-type solutions, respectively.}
\end{figure}

The Kasner solutions (for $\Lambda >0$) get deformed and the functions $M$, $L$
become periodic in $x$ asymptotically. 
These are deformations of the periodic solution (\ref{trigonometric})
with $C\neq 0$. 
This is illustrated in Fig.\ref{fig10}.
The value of $C$ for these solutions is negative. The function
$M$ oscillates around a mean value given by 
$M_{mv}= (\frac{-3 C}{2 \Lambda})^{1/5}$
and stays strictly positive, the solution
is therefore regular. The period of the solution depends weakly on $\kappa$.
In the limit, $\Lambda \to 0$ the periodic solutions tend  to the cigar-like
solutions. 
 As mentioned above, we notice that the values of the constants
$\kappa$ and $\Lambda$ can be chosen in such a way that 
$2\kappa - \Lambda$ becomes arbitrarily small. This
seems compatible with the domain of parameters leading to periodic 
solutions. The corresponding Planck mass determined through 
(\ref{plmas})
can therefore be made arbitrarily large.

A sketch of the pattern of the solutions is proposed
in Fig.\ref{fig11}.
The fact that we obtain periodic solutions
constitutes a natural framework to define finite volume brane world models.
With view to the discussion on the effective 4-dimensional
action, we could imagine to put branes with ``large''
effective gravitational coupling at $x=x_k$, $k=1,2,3,...$, 
where $x_k$ corresponds to the position of the maxima of $M(x)$.
Correspondingly, branes with ``small'' effective gravitational
coupling could be put at $x=\tilde{x}_k$, $k=1,2,3,..$, where
$\tilde{x}_k$ are the positions of the minima of $M(x)$.
Though there is no ``warping'' in the sense of Randall and Sundrum \cite{rs1,rs2},
nevertheless gravity looks stronger on the branes positioned at the maxima
of $M(x)$ in contrast to gravity on the branes positioned at the minima.
Note that at the positions of these branes, the metric function $L(x)$
vanishes, i.e. $L(x_k)=L(\tilde{x}_k)=0$.
Thus the cylindrical geometry of the extra dimensional space 
shrinks to a point and the whole space-time geometry becomes effectively
5-dimensional.

\section{Summary and Discussion}
We have studied static as well as inflating brane solutions in a 6-dimensional
brane world scenario with an abelian string residing in the two extra dimensions.
The pattern of the solutions is very rich, leading
to many different types of behaviours of the gravitating fields  
(depending on the values of $\gamma$, $\kappa$ and $\Lambda$)
outside the core of te string. We find it convenient to summarize the
possiblilites as follows
\begin{itemize}
\item String solutions (S) with $M \to 1$, $L \to \infty$ for $x\to \infty$,
\item Melvin solutions (M) with $M \to \infty$ , $L \to 0$ for $x\to \infty$,
\item Inverted string solutions (IS) with
$M \to 0$ , $L \to -\infty$ at some finite value of $x$
       ($L(x_1)=M'(x_1)=0$ for some intermediate value of $x=x_1$),
\item Kasner solutions (K) with $M\to 0$, $L \to \infty$
at some finite value of
$x$, 
\item Cigar-type solutions  (C) with
$M/x \to M_0$, $L \to L_0$ for $x \to \infty$,
\item Periodic solutions (Per) with $M$, $L$ periodic in the bulk coordinate $x$.
\end{itemize}
We point out again  here that the different names used to 
characterize the solutions
is an adaptation of the terminology used e.g.
in \cite{gstring} to the present context.

Evaluating the different curvature invariants of the metric we find that
the Kasner and inverted string solutions possess a physical
singularity at the value of $x$ at which the function $M$ vanishes. Note that
this excludes the inverted string solutions for static branes in an asymptotically
flat bulk, where $M$ possesses no zero. These solutions were  shown to be
regular \cite{bh2}.
All the other space-times are without physical singularities.
If the symmetry breaking scale is below a critical below (given by the
higher-dimensional Planck scale) corresponding to $\gamma < 2$,
the space-time containing a static brane is singularity-free for an asymptotically flat
or Anti-de-Sitter bulk, however has singularities for an asymptotically de Sitter bulk (positive
bulk cosmological constant). Inflating branes lead to a cigar-like universe for vanishing
bulk cosmological constant  at a symmetry breaking scale below a specific
value that depends on the value of the brane cosmological constant. Next to these, also Melvin solutions
exist, which again are singularity-free.

If the symmetry breaking scale is above a critical value, 
the space-time is singular for static branes
in an asymptotically flat bulk ($\Lambda=0$).
If we replace the static brane by an inflating brane,
the space-time still possesses a singularity. 
Equally, the introduction
of a positive cosmological constant doesn't remove the singularity of the space-time.

If both the bulk cosmological constant and the brane cosmological constant
are non-vanishing, regular space-times are possible, 
namely periodic solutions with non-vanishing minima
of the metric functions. 
These  appear if in the space-time with $\Lambda > 0$ (Kasner solutions)
the static brane is replaced by an inflating brane.
If one puts branes at the respective minima and maxima of these
periodic solutions, a hierarchy between the effective 4-dimensional
gravitational coupling is possible.

As a final remark let us relate our periodic solutions to recently investigated solutions in
higher dimensional space-times,
namely  ``Non-uniform black strings''  (see e.g. \cite{wiseman}).
Indeed, referring to Fig.1  of \cite{wiseman}, the solution in the extra
dimension could be continued  periodically by gluing several copies of the
solution. The structure in the extra dimension then becomes kind of periodic.
The crucial difference, of course, is that in the present paper
we don't have black holes extended into extra dimensions, but an
inflating brane (the 3-brane) with two extra dimensions. However,
again, we find a periodicity of the solution in the extra dimension,
in our case in the $x$-direction.
The periodic solutions presented here could thus be seen as
``non-uniform De Sitter-vacuua''.\\
\\
{\bf Acknowledgements} \\
YB is grateful to the
Belgian FNRS for financial support.

\newpage
\newpage

\newpage
\newpage

\newpage
\newpage

\newpage

\newpage

\newpage

\newpage

\newpage
\newpage

\end{document}